\documentstyle[preprint,prl,aps]{revtex}

\newcommand{\be}{\begin{equation}}
\newcommand{\ee}{\end{equation}}
\newcommand{\BE}{\begin{eqnarray}}
\newcommand{\EE}{\end{eqnarray}}
\newcommand{\BEn}{\begin{eqnarray*}}
\newcommand{\EEn}{\end{eqnarray*}}
\newcommand{\barr}{\begin{array}} 
\newcommand{\earr}{\end{array}}
\newcommand{\bit}{\begin{itemize}}      
\newcommand{\eit}{\end{itemize}}
\newcommand{\bfl}{\begin{flusleft}}
\newcommand{\efl}{\end{flusleft}}
\newcommand{\bfr}{\begin{flushright}}
\newcommand{\efr}{\end{flushright}}

\newcommand{\bc}{\begin{center}}
\newcommand{\ec}{\end{center}}

\newcommand{\ben}{\begin{enumerate}}    
\newcommand{\een}{\end{enumerate}}

\newcommand{\sol}{\overline \sigma}

\newcommand{\pr}{\protect}

\newcommand{\de}{\partial}
\newcommand{\eps}{\varepsilon}
\newcommand{\Dm}{\frac{\Delta_1}{2}}

\begin{document}
\draft

\title{Analytic estimate of the maximum Lyapunov exponent \\
in coupled-map lattices}
\author{F. Cecconi$^{1,2,\dagger},$ and A. Politi$^{3,2}$}
\address{  
$(1)$ Dipartimento di Fisica, Universit\`a di Firenze \\
$(2)$ INFM, Unit\`a di Firenze\\
$(3)$ Istituto Nazionale di Ottica, Firenze, Italy\\
$\dagger$ Also with INFN Sezione di Firenze}
\date{}

\maketitle
\begin{abstract}
In this work we present a theoretical and numerical study of the behaviour
of the maximum Lyapunov exponent for a generic coupled-map-lattice in
the weak-coupling regime. We explain the observed results by introducing a 
suitable continuous-time formulation of the tangent dynamics. The first 
general result is that the deviation of the Lyapunov exponent from the 
uncoupled-limit limit is function of a single scaling parameter which,
in the case of strictly positive multipliers, is the ratio of the coupling 
strength with the variance of local multipliers. Moreover, we find an 
approximate analytic expression for the Lyapunov exponent by mapping the 
problem onto the evolution of a chain of nonlinear Langevin equations, which 
are eventually reduced to a single stochastic equation. The probability
distribution of this dynamical equation provides an excellent description 
for the behaviour of the Lyapunov exponent. Furthermore, multipliers 
with random signs are considered as well, finding that the Lyapunov 
exponent still depends on a single scaling parameter, which however has a
different expression. 
\end{abstract}
\vskip 1.cm
Short title: Maximum Lyapunov exponent in coupled-map lattices

\pacs{05.45.-a}


\newpage
\narrowtext

\section{Introduction}

Coupled-map lattices (CML) represent an interesting class of models for
the investigation of several spatio-temporal phenomena, ranging from pattern
formation to synchornization and to spatio-temporal chaos. Even though the
discreteness of both space and time variables makes CML more amenable to
numerical simulations than partial differential equations, the development
of analytical techniques remains a difficult task. As usual, when dealing
with hard problems, it is convenient to start from the identification of some
relatively simple limit and thereby developing a perturbative approach. In the 
case of lattice dynamics, there are two opposite limits that are worth being
investigated: the weak- and strong-coupling regime. In the former case, one
can use all the knowledge acquired about low-dymensional systems to predict
the dynamical properties when spatial directions are added. In this spirit,
general theorems have been formulated and proved both about the structure of
the invariant measure\cite{bunimovich} and about the existence of travelling
localized excitations\cite{mackay}. The weak-coupling limit is interesting
also in connections with the synchronization of chaotic attractors, a
problem that can be effectively studied by looking at the behaviour of the
Lyapunov exponents \cite{pikoww}. In the opposite limit, one expects a
slow spatial dependence and, correspondingly, the existence of a few,
dynamically active, degrees of freedom. This is the spirit that has led to
studying different truncations of partial differential equations.

In this paper, we shall consider the most common indicator of chaos, the
maximum Lyapunov exponent (MLE) in a lattice of coupled maps. It is known
that a (weak) spatial coupling has a threefold consequence on the MLE:
{\it i)} it naturally modifies the invariant measure; {\it ii)} it induces
correlations among the local multipliers by coupling neighbouring sites;
{\it iii)} it modifies the dynamics of perturbations by inducing a coupling
in tangent space as well. The last phenomenon is the most interesting one,
since it leads to counterintuitive effects like an increase of the MLE even
in the presence of a stabilizing coupling.

In some previous papers \cite{LPR,CP}, this problem has been addressed with
a specific interest for the scaling behaviour of the MLE. However, the
various approaches implemented so far have not been able to go beyond a
qualitative explanation of the dependence on the coupling strength.
Here, instead, we aim at presenting a fully quantitative, though approximate,
treatment for the MLE in the small coupling regime. We restrict our analysis
to the usual diffusive coupling scheme, but we are confident that the
present approach can be effectively adapted to different (short-range)
interaction schemes. More specifically, we refer to the dynamical equation
\BE
x_i(t+1) &=& f (y_i(t+1)) \nonumber \\
y_i(t+1) &=& \eps x_{i-1}(t) + (1-2\eps) x_i(t) + \eps x_{i+1}(t) .
\label{eq:cml}
\EE
The corresponding evolution equation in the tangent space is
\be
u_i(t+1) =  m_i(t)
\bigg\{\eps u_{i-1}(t) + (1-2\eps) u_i(t) + \eps u_{i+1}(t) \bigg\},
\label{eq:cmlt}
\ee
where $m_i(t) \equiv f'(y_i(t+1))$.
From Eq.~(\ref{eq:cmlt}), we see that the last of the above mentioned
three effects of the spatial coupling can be isolated and studied separately
from the other ones. It is sufficient to assume that the distribution of 
multipliers is independent of the the coupling strength $\eps$
and that the $m_i(t)$'s are $\delta$-correlated both in space and time.
These approximations are equivalent to introducing a random-matrix
approximation in full analogy with Refs.~\cite{LPR,CP}.

The first author who investigated the effect of a small coupling on a chaotic
dynamics was Daido \cite{daido}, who studied numerically two coupled maps as
well as two continuous-time chaotic oscillators. He also attempted a
combined analytical and numerical study to explain the observed behaviour,
without however being able to go beyond the prediction of the scaling
behaviour.

Later on, the problem of (infinitely many) coupled maps was considered
in Ref.~\cite{LPR}, by exploiting the analogy with the statistical
mechanics of directed polymers in random media. Indeed, Eq.~(\ref{eq:cmlt})
can be also read as the recursive equation for the partition function
``$u_i(t)$'' of a polymer of length $t$ which grows by adding each new
monomer no farther than one site from the last one. A possibly relevant
difference between the two problems comes from the ``Boltzmann weight''
$m_i$ which is necessarily positive in the polymer case (being a
probability), while it can be negative in a CML ($m_i$ being the derivative
of the map $f$). This is a first issue that makes the problem
(\ref{eq:cmlt}) more difficult to be treated and it is the reason why
previous studies have been restricted to the case of strictly positive
multipliers \cite{LPR,CP}. In fact, without entering the mathematical 
treatment, one can
see that if $m_i$ can be either positive or negative, $u_i(t)$ is no longer
positive definite and partial cancellations can occur in the iteration of
the recursive relation.

The efforts made in \cite{LPR} to estimate the MLE consisted in developing a
mean field analysis on the basis of the equivalence between the MLE in CML
and the free-energy in directed polymers. By thus using the approaches
developed in \cite{DS88,CD89}, it was possible to show that the spatial
coupling induces an increase of the MLE from the ``quenched'' average
$\Lambda_0 = \langle \log m_i \rangle$ (corresponding to the absence of
a spatial coupling) towards the ``annealed'' average
$\Lambda_0 = \ln \langle m_i \rangle$, holding above some critical coupling
value. While some features of this scenario were qualitatively confirmed by
the numerical simulations (as, e.g., the increase of the MLE), no evidence
of the phase transition was actually found. A perturbative technique,
developed later on to improve the previous estimates \cite{CP} has shown 
that the transition point slowly shifts towards larger coupling values, 
possibly disappearing in the asymptotic limit. Nevertheless, the extremely 
slow convergence of the estimates of the MLE to the values numerically 
observed, makes a general
implementation of this approach not very appealing. Moreover, we should also
recall that the analogy with directed polymers does not even allow an exact
prediction of the scaling behaviour of the MLE, in so far as it indicates
the existence of an additional, extremely weak, dependence on the coupling
strength which seems to be absent in the outcome of direct numerical
simulations.

It is also worth recalling the analogy between the behaviour of the MLE
and the evolution of rough interfaces. By interpreting the logarithm of (the
amplitude of) the perturbation as the height of an abstract interface,
the Lyapunov exponent becomes equivalent to the velocity of one such
interface \cite{PiKu,PiPo}. However, this analogy is of no utility in the
present case, since the deviation of the MLE from the uncoupled limit cannot
be determined by studying the corresponding Kardar-Parisi-Zhang (KPZ)
equation as already remarked in \cite{CP}. In a sense, the MLE corrections 
are connected to non-trivial deviations from a KPZ behaviour over short 
temporal and spatial scales.

In the present paper we derive approximate but analytical expression for the
MLE, by mapping the original problem onto that of a chain of
continuous-time, nonlinear Langevin equations. Such a set of equation is
then reduced to a single stochastic equation whose solution yields an
expression for the MLE that is in good agreement with direct numericals
simulations. As the whole approach does not make use of the specific
structure of the initial equations, we are rather confident that it can be
repeated for other types of couplings, the only difference being presumably
the structure of the deterministic force in the final stochastic equation.
Moreover, we would like to point out that the mapping onto continuous-time
equations indicates that the initial discreteness of the time variable is
not a distinctive feature and we can imagine that a similar approach works
also in the case of coupled chaotic oscillators. Our hope is reinforced by
the observation that some equations obtained in the present framework can be
derived also in the case of two weakly coupled differential equations
\cite{pikoco}.

The paper is organized as follows. In Sec. II, we briefly introduce the
problem and some notations. In Sec. III we discuss the properties of tangent
dynamics in the case of strictly positive multipliers: the analytical
treatment is followed by a comparison with the numerical results.  In
Sect.~IV we extend the analytical treatment to the case of multipliers with
random signs. Finally, in Sect.~V we summarize the main results and comment
about the still open problems. The two appendices are devoted to the
small-noise limit in the case of strictly positive multipliers and,
respectively, to the small coupling regime in the general case of random
signs.

\section{Preliminary treatment}

In this section, we formulate the problem of computing the MLE under
the assumption of a small coupling strength. The first step consists in
introducing the ratio between the perturbation amplitude in two consecutive
site,
\be
R_j(t) \equiv u_{j}/u_{j-1}
\label{rappo}
\ee
which allows writing Eq.~(\ref{eq:cmlt}) as
\be
\ln \left \vert \frac{u_i(t+1)}{u_i(t)} \right \vert =
 \ln \vert m_i(t)\vert + \ln \left \vert  1 - 2\eps + \eps R_{i+1}(t) +
\frac{\eps}{R_i(t)} \right \vert ,
\label{eq:tgratio}
\ee
where the average of the l.h.s. is nothing but the MLE $\Lambda(\eps)$,
while the r.h.s. is naturally expressed as the sume of the 0-coupling value 
plus the correction term induced by spatial interactions. 

A more convenient way of writing the Lyapunov exponent is obtained by
transforming the smallness parameter as
\be
\gamma \equiv \eps/(1-2\eps) , 
\ee
which leads to 
\be
\Lambda(\eps) \;=\; \bigg\langle \ln \left \vert 
\frac{u_i(t+1)}{u_i(t)} \right \vert \bigg \rangle 
\; =\; \Lambda(0) + \ln(1-2\eps) + \delta\Lambda(\gamma) ,
\label{eq:lyap}
\ee
where the Lyapunov correction is split into two parts, a
multiplier-independent term and a non-trivial contribution
\be
\delta\Lambda(\gamma) = \bigg\langle \ln \bigg|1 + \gamma R_{i+1} +
\frac{\gamma}{R_i} \bigg| \bigg\rangle ,
\label{eq:shift}
\ee
where $\langle\cdot\rangle$ represents the time-average along the trajectory 
generated by Eq.(\ref{eq:cml}) or, equivalently, the ensemble-average if 
ergodicity holds.

Eq.~(\ref{eq:shift}) tells us that the determination of the MLE requires
the prior knowledge of the invariant measure of the stochastic process 
$R_i(t)$ on two consecutive sites. 
By expanding the logarithm in powers of $\gamma$, the 
correction to the MLE can be expressed in terms of all the momenta of the 
probability distribution,
\be
\delta\Lambda(\gamma) = - \sum_{j=1}^{\infty}\; \sum_{m=1}^{j}
 \frac{(j-1)!(-\gamma)^j}{m!(j-m)!}\;\; \left \langle 
   R_{i+1}^m R_{i}^{j-m} \right \rangle .
\label{eq:expan}
\ee
It is, therefore, necessary to formulate the dynamical equation in terms
of the variable $R_i$. This can be easily done from
Eqs.~(\ref{rappo},\ref{eq:tgratio})
\be
R_i(t+1) \; = \; \mu_i(t)\; \frac{R_i(t)+\gamma R_i(t)\; R_{i+1}(t)+\gamma}
{1 + \gamma R_i(t) + \gamma /R_{i-1}(t)}. 
\label{eq:Rmap}
\ee
where we have introduced the stochastic process
\be
\mu_i(t) \equiv m_{i}(t)/m_{i-1}(t)
\ee
whose geometric average is equal to 1, $\mu_i$ being the ratio of two i.i.d.
processes.

\section{Positive multipliers}

We first consider strictly positive multipliers, while the more general case 
is addressed in the following section. In fact, while the two cases require a
somehow different treatment, a discussion of the former problem allows
introducing several tools that turn out to be useful also in the latter
context.

\subsection{Theory}

If the $\mu_i$'s are positive definite, the ratios $R_i$'s remain positive 
whenever initialized as such. This allows introducing the variable 
$w_i = \ln R_i$ without the need of dealing with the sign of $R_i$.
The introduction of $w_i$ is convenient in that it transforms the original 
problem into a stochastic process with additive rather than multiplicative 
noise. With no restriction other than the positiveness of the multipliers,
we obtain
\BE
w_{i}(t+1) - w_i(t) & = & \ln \mu_i(t) + 
\ln\bigg\{1 + \gamma e^{w_{i+1}(t)} + \gamma e^{-w_{i}(t)} \bigg\}
-\nonumber \\
& &\ln\bigg\{1+ \gamma e^{w_{i}(t)} + \gamma e^{-w_{i-1}(t)} \bigg\}  .
\label{eq:wmap}
\EE
In the limit of small $\gamma$, the arguments of the logarithms differ 
slightly from 1, provided that $w_i$ does not deviate too much from 0. Under 
this approximation (that can be checked a posteriori), we can expand the
logarithms, retaining the leading terms and thereby introducing a 
continuous-time representation (because of the smallness of the 
deterministic variations). One obtains
\be
\dot{w_i} =  -2\gamma \sinh(w_i) + \gamma\bigg(e^{w_{i+1}} - e^{-w_{i-1}}\bigg)
+ \xi_i(t) 
\label{eq:wdot}
\ee
where $\xi_i \equiv \ln \mu_i$, has zero average and correlation function,
\be
\langle \xi_i(t) \xi_j(t')\rangle = 4 \sigma^2 \delta(t-t')
(\delta_{i,j} - {1\over2}\delta_{i\pm 1,j})
\label{eq:corr}
\ee
where $\sigma^2$ is the variance of the logarithm of the multiplier $m_i$ and 
there is a factor 4 (instead of the usual 2) since the noise term is 
the ``sum'' of two i.i.d. processes ($\xi_i = \ln m_i - \ln m_{i-1}$). 
Spatial correlations of the stochastic forces are all zero except those of
neighbouring sites, a feature induced by the very definition of $\xi_{i}$ as 
the sum of two processes in neighbouring sites.

It is important to notice that the only parameter of the sequence of
multipliers which eventually contributes to the correction of the MLE is
precisely the amplitude of the fluctuations. 

Eq.~(\ref{eq:wdot}) represents a chain of coupled Langevin equations
describing the evolution of interacting ``particles''. Even without solving
the model, it is possible to realize that only one parameter suffices to
describe the scaling behaviour of the Lyapunov exponent, the rescaled
smallness parameter
\be
 g  =  \gamma/\sigma^2 ,
\label{defg}
\ee
which can also be interpreted as the inverse effective diffusion constant.
In fact, the factor $\gamma$ in front of the deterministic forces can be
eliminated by properly scaling the time units.

In the next sub-section, we investigate numerically the validity of this
prediction by computing the Lyapunov exponent for different values of
$\sigma$ and $\gamma$ and thereby checking whether the data collapses onto
a single curve.

The evaluation of $\delta\Lambda(\gamma)$ requires finding the invariant
measure for the whole set of stochastic equations (\ref{eq:wdot}). This is
still a difficult problem, since the deterministic forces do not follow from
a potential and, therefore, the corresponding Fokker-Planck equation cannot
be straightforwardly solved. In particular, it is interesting to notice
that, although we are working in the limit of weakly coupled maps, the
``particles'' are not weakly interacting. This is the most serious
difficulty towards a perturbative treatment of the problem. Nevertheless, we
are going to see that this infinite set of stochastic equations can be
effectively approximated by a single Langevin equation.

Formally speaking, the probability distribution $P_L(w_1,...,w_L\hbox{;}t)$
satisfies the Fokker-Planck equation
\be
\frac{\de P_L}{\de t} =
- \sum_{i=1}^{L} \frac{\de}{\de w_i} \bigg\{ F(w_i) P_L +
\Phi(w_{i+1},w_{i-1}) P_L \bigg\} +
\frac{1}{2}\sum_{i,j=1}^{L} {\cal D}_{i,j}\frac{\de^2 P_L}{\de w_i\de w_j}
\label{eq:FPw}
\ee
where
\be
F(w_i) = -2\gamma \sinh(w_i)
\label{force0}
\ee
is the single-particle force (see Eq.~(\ref{eq:wdot}),
$\Phi(w_{i+1},w_{i-1}) = -2\gamma ({\rm e}^{w_{i+1}} - {\rm e}^{-w_{i-1}})$
is the coupling term and ${\cal D}_{i,j}$ are the diffusion coefficients
as from (\ref{eq:corr}).

Let us now introduce the single-particle distribution $P(w_i)$ by 
integrating over all the other $L-1$ degrees of freedom
$$
P(w_i,t) = \int \prod_{j\ne i} dw_j\;P_L(w_1,...,w_L;t)\,.
$$
The corresponding equation can be directly derived from Eq.~(\ref{eq:FPw}),
\BE
\label{eq:FPw1}
\frac{\de P}{\de t} & = &  -\frac{\de}{\de w_i}\{F(w_i) P\} +
2\sigma^2 \frac{\de^2 P}{\de w_i^2} + \\  \nonumber
                      &   &  -\gamma \frac{\de}{\de w_i}
\bigg\{\int dw_{i+1} P_2(w_i,w_{i+1}) {\rm e}^{w_{i+1}} -
       \int dw_{i-1} P_2(w_{i-1},w_i) {\rm e}^{-w_{i-1}} \bigg\}
\EE
Eq.~(\ref{eq:FPw1}) is not closed since it involves the unknown two-body
distribution $P_2$. The above is indeed the first of a hierarchy of
equations involving probability distributions of increasing order. The
simplest approximation to close the system consists in truncating
the hierarchy at the lowest level by assuming a perfect factorization,
$P_2(x,y) = P(x)P(y)$. This is the standard mean-field approach which leads
to the closed Fokker-Planck equation
\be
\frac{\de P}{\de t} = -\frac{\de}{\de w}\{F(w) P\} +
2\sigma^2 \frac{\de^2 P}{\de w^2} ,
\label{eq:FP}
\ee
since the symmetry of the distribution ($P(w) = P(-w)$) implies that the
coupling terms cancel each other.

The Fokker-Planck equation (\ref{eq:FP}) corresponds to the single-particle
Langevin equation
\be
\dot{w} =  F(w) + \xi(t)  ,
\label{eq:wappr}
\ee
where we have dropped the by now irrelevant spatial dependence.

It is instructive to test the validity of the above approximation in the
limit of large $g$ values, when the forces can be linearized and an
analytic solution can be obtained for the entire chain of Langeving
equations. For the sake of readability, the discussion of this technical
problem is presented separately in the first Appendix. The main result of
this analysis is that the approximation (\ref{eq:wappr}) is exact!
The stationary probability distribution of Eq.~(\ref{eq:FP}) is equal to
the projection of the many-particle distribution for the whole chain.
However, one cannot expect that the same holds true also for finite
$g$ values, when nonlinearities come into play.

Let us therefore discuss a possible line of thought to go beyond the
approximation (\ref{eq:wappr}). It is first important to recognize
the special nature of the noise $\xi$, which keeps $W(t) = \sum_i w_i(t)$
constantly equal to zero. Indeed, $W$ is nothing but the logarithm of the
ratio between the amplitude of the perturbation in the first and the last
site. If periodic boundary conditions are assumed, as we do,
this implies that $W(t)=0$. A simple way to satisfy this constraint is
to assume $w_i = -w_{i-1}$. By substituting this extreme assumption into
Eq.~(\ref{eq:wdot}), we obtain again a Langevin equation with the same
force as in Eq.~(\ref{force0}) but a different factor. It is therefore
reasonable to conjecture that, in general, the effective force strength
depends on $g$, thus writing (in rescaled time-units)
\be
F(w)  = - 2\alpha (g) \sinh(w)
\label{force}
\ee
where $\alpha$ is a renormalization factor to be determined with some
self-consistency argument. As we have not found any sensible way to do so,
we limit ourselves to investigate numerically its possible dependence on $g$.

The stationary distribution $w$ is obtained by solving the
Fokker-Planck equation (\ref{eq:FP}),
\be
P(w) = {\cal C} \exp[-\alpha \gamma/\sigma^2\; \cosh(w)],
\label{eq:pofw}
\ee
where the normalization constant ${\cal C} = 2 K_0(\alpha \gamma/\sigma^2)$ is 
expressed in terms of the zeroth-order modified Bessel function 
\cite{GrasRyz,Abr}
$$
K_{\nu}(x) = \int_{-\infty}^{\infty} dt \exp\{-x \cosh(t) \pm\nu t\} .
$$
By substituting the definition of $w$ ($w = \ln R$) in Eq.~(\ref{eq:expan}),
and assuming $\langle w_i w_{i-1}\rangle =0$ (an hypothesis accurately
confirmed by direct numerical simulations), we finally obtain
\be
\delta\Lambda(\gamma) = -\sum_{j=1}^{\infty}\; \sum_{m=1}^{j}
\frac{(j-1)!(-\gamma)^j}{m!(j-m)!}\; \frac{K_{m}(2g) K_{j-m}(2g)}{K_0(2g)^2}
\label{eq:teocle}
\ee
As $\gamma$ is assumed to be small, we can safely retain only the first two
terms of the series (\ref{eq:teocle}), obtaining
\be
\delta\Lambda(\gamma) = 2 \gamma\; \frac{K_1(2g)}{K_0(2g)}
\label{eq:cle1}
\ee
which represents the (approximate) perturbative expression for the
correction to the MLE in the limit of small coupling.

First of all, it is instructive to investigate the limit $g\ll 1$, by
using the asymptotic expression of the functions $K_{\nu}(y)$. Since
\be 
K_{0}(y) \sim -\ln (y/2) \quad\quad, \quad\quad K_1(y) \sim {1\over y}  ,
\ee
we find that
\be
\delta \Lambda(\gamma) \quad \sim \quad
  \frac{\sigma^2}{\ln(1/g)}
\label{eq:scale}
\ee
This equation represents a relevant improvement over the previous
results. First of all, it is in agreement with numerical simulations
which do not give evidence of a $\ln |\ln g|$ correction in the numerator,
as instead predicted by the statistical-mechanics treatment based on the
analogy with directed polymers \cite{CP}.

A second and more important remark concerns the dependence on the ``noise'' 
strength which is explicitely determined. Previously it was only clear that
the correction to the MLE must vanish if there is no multifractality
(no multiplier fluctuations) but the dependence on $\sigma$ was not known.

\subsection{Numerical results}

The theoretical analysis performed in the previous section is mainly
based on a perturbative approach. Moreover, it involves a nontrivial
transformation of a set of coupled Langevin equations to a single 
effective Langevin equation in a limit where the coupling is not negligible. 
Therefore, a comparison of the theoretical predictions with direct numerical
simulations is worth especially to check the validity of the dynamical
mean-field approximation that is behind this last step.

We have decided to test the thereotical results by using two different
probability densities: A) uniform distribution of multipliers
$m_i$ within the interval $[{\rm e}^a(1-\Delta_1/2),{\rm e}^a(1+\Delta_1/2)]$;
B) uniform distribution of the logarithms of the $m_i$ within
$[a-\Delta_2/2,a+\Delta_2/2]$. The corresponding Lyapunov exponents in the
uncoupled limit ($\eps=0$) are
$$
\Lambda_A(0) = a -1 +
   {1\over\Delta_1}\bigg\{\bigg(1+\Dm\bigg)\ln\bigg(1+\Dm\bigg) -
   \bigg(1-\Dm\bigg)\ln\bigg(1-\Dm\bigg)\bigg\}
$$
$$
\Lambda_B(0) = a, 
$$
while the variances of $\ln m_i(t)$ are
\be
\sigma_A^2 =  1 - \frac{1}{\Delta_1^2} \bigg(1 -\frac{\Delta_1^2}{4}\bigg)
   \ln^2 \left (\frac{1+\Delta_1/2}{1-\Delta_1/2}\right) \nonumber
\ee
\be
\sigma_B^2 = \frac{\Delta_2^2}{12} . \nonumber
\ee
We start from testing the predictions for the shape of the probability
distribution of $w$. Two meaningful examples are reported in 
Fig.~\ref{fig:histow}, where the open circles refer to the histograms,
while the solid line is the theoretical result (Eq.~(\ref{eq:pofw}) with 
$\alpha$ set equal to 1).
Let us first comment about the qualitative shape of the distribution. In the 
limit of large $g$, the noise is almost negligible and therefore, the phase 
point is expected not to deviate significantly from the stable fixed point 
$w=0$. It is therefore possible to linearize the equation, finding a Gaussian 
distribution. This is precisely the message contained in 
Fig.~\ref{fig:histow}a, which refers to $g=2.4$. In the limit of small $g$ 
instead, it is the force that can be neglected except when the deviations are 
large. Since the attracting force grows very rapidly (exponentially), it 
makes sense to replace the corresponding
potential with a flat well with infinitely high barriers placed where
the deterministic force is of the same order as the stochastic one.
In this picture, one expect that the probability distribution is just
a uniform distribution in a finite interval (this is the kind of argument
introduced in \cite{LPR} to predict the scaling behaviour in this regime). 
This scenario can be qualitatively recognized in Fig.~\ref{fig:histow}b, 
which refers to $g=0.021$.

Next, let us comment about the quantitative agreement between the
theoretical expectations and the numerical findings. In Fig.~\ref{fig:histow}a,
there is an almost perfect agreement. This is a first encouraging result,
since it indicates that even for not too large a value of $g$, the effective
force strength remains equal to $2\gamma$ with no renormalization.
Some deviations are instead observed in Fig.~\ref{fig:histow}b, where
the theoretical curve is slightly more peaked, suggesting that the effective
force is smaller than expected. A simple way to determine $\alpha$ is by 
fitting the numerical data. In the first case we find that $\alpha = 0.97$ 
confirming the first qualitative impression; in the second case
$\alpha = 0.75$ indicating stronger but not significative deviations from 
the linear regime. The curve corresponding to the fitted value of $\alpha$ 
is reported in Fig.~\ref{fig:histow}b as a dashed line. The residual, small, 
deviations with respect to the numerical data indicate that the reduction of
the original model to a single stochastic equation with the effective force 
(\ref{force}) is indeed a strategy that is worth being pursued.

All the other cases that we have tested reveal the same scenario. Altogether,
we can summarize stating that the major discrepancy between numerics and
theoretical prediction is contained in the renormalization factor $\alpha$ 
which is assumed to be equal to 1 in our treatment but turns out to depend
on $g$; nevertheless it is never smaller than $0.7$. 

Since the aim of the present paper is to study the corrections to the MLE
induced by the spatial coupling, let us now discuss this issue. In order to 
assess the quality of our theoretical predictions, we performed numerical 
iterations of Eq.~(\ref{eq:cmlt}) computing the MLE with the well 
known algorithm \cite{galga}. Simulations have been carried out on 500-site 
lattices, imposing periodic boundary conditions. In every simulation the 
first 500 iterations have been discarded to avoid any bias effect due to 
initial conditions. Tests made with different lattice lengths indicate that 
finite-size effects are always much smaller than the deviation from the 
theoretical predictions.

The first nice result is provided by Fig.~\ref{fig:scale}a. The data is 
plotted in order to emphasize the existence of only one relevant parameter,
$g$. Indeed, the good data collapse (all data align along the same curve 
irrespective of the value of $\gamma$, $\sigma$ or the type of probability 
distribution) represents a further confirmation of our theoretical analysis.
Moreover, the nice agreement of the numerical data with the theoretical
expression (\ref{eq:cle1}) (see the solid line in Fig.~\ref{fig:scale}a)
over a wide range of values of the effective coupling testifies to the
accuracy of the approximations introduced in the first part of this section.

However, there is a better way to emphasize the differences between theory and
numerics. In fact, the limit $g \to \infty$ corresponds to negligible noise, 
i.e. to a regime where the MLE Lyapunov exponent is unaffected by the presence
of spatial coupling as shown in \cite{iprt}. It is therefore convenient to
look at the behaviour of the whole deviation of the Lyapunov exponent 
$\Delta \Lambda = \delta \Lambda - 2\gamma$, which again exhibits the same
scaling behaviour as seen by dividing this expression by $\sigma^2$
and using Eq.~(\ref{eq:scale})
\be
 \frac{\Delta \Lambda}{\sigma^2} = 2 g \left (\frac{K_1(2g)}{K_0(2g)} - 1
 \right ) .
\label{eq:globly}
\ee
The data plotted this way are reported in Fig.~\ref{fig:scale}b. We clearly
see that the trivial correction term $-2\gamma$ cancels almost exactly the
growth exhibited by $\delta \Lambda$ for large $g$-values allowing for a
more stringent test of the theoretical prediction. We can now see that
the absolute difference is not larger than 0.05 and it could be greatly
reduced by suitably shifting the theoretical curve (i.e. by rescaling
$g$ by approximately a factor 2) as shown by looking at the dashed curve.
However, in the absence of theoretical arguments, this observation cannot
be considered more than a hint for future considerations.

\section{The general case}

\subsection{Theory}
In this section, we account for sign fluctuations as well. In order to
keep the theoretical treatment as simple as possible, we shall assume that
the sign is a $\delta$-correlated stochastic process independent of the
modulus, so that the average factorizes as
$$
\langle m_i \rangle = (p-q) \langle |m_i| \rangle ,
$$  
where $p$ ($q$) is the probability that $m_i$ is positive (negative). The major
difference with respect to the previous case is that the ratio $R_i$ can also 
assume negative as well as positive values. It is therefore more convenient to 
work with $R_i$ instead of introducing its logarithm which would 
require introducing absolute values and thus two different variables to account
for the dynamics in the positive as well as in the negative range of $R_i$ 
values.

We have learned in the previous section that neglecting the coupling with
the neighbouring sites provides a good approximation of the probability
distribution and thereby of the MLE. Let us therefore neglect the dependence
on $R_{i+1}$ and $R_{i-1}$ in Eq.~(\ref{eq:Rmap}). As a result, we obtain the 
one dimensional mapping
\be
R(t+1) \; = \; \mu(t)\; \frac{R(t) + \gamma}{1 + \gamma R(t)} ,
\label{eq:Rmaps}
\ee
where, for the sake of simplicity, we have removed the now irrelevant site 
dependence. In the continuous-time approximation, the above would be exactly 
the equation that has allowed an approximate analytical treatment in the case 
of positive signs. Mapping (\ref{eq:Rmaps}) possesses
two remarkable symmetry properties: the evolution is invariant under the
transformation $R \to 1/R$, since the stochastic process $\mu$ turns out to
be invariant under the same transformation $\mu \to 1/\mu$ (it is sufficient
to look at its definition). This is the same symmetry as that one discovered
in the previous section when we have found that the potential $V(w)$ is an
even function of $w$. As a consequence, we can restrict our analysis to the 
interval $[-1,1]$.

The second symmetry has much more serious implications. We can see that
the mapping (\ref{eq:Rmaps}) is also invariant under time-reversal. More
precisely, if we express $R(t)$ as a function of $R(t+1)$ we find the
same functional form of mapping (\ref{eq:Rmaps}), provided that the changes 
of variables $S: R(t+1) \to - R(t+1)/\mu(t)$ and $\mu \to 1/\mu$ are
introduced as well. As the transformation $S$ is an involution 
(i.e., $S^2 = Id$), we can state that mapping (\ref{eq:Rmaps}) is 
invariant under time-reversal.  Therefore, we seem to be in the presence of a 
contradiction, as this property holds also for strictly positive multipliers,
when it is clear that there is an attractor (the point $w=0$, i.e. $R=1$),
since time-reversibility hints at a lack of attractors! Indeed, there is 
no contradiction, since time-reversal simmetry is broken in the case of
strictly positive multipliers. In fact, invariance of the mapping implies 
only that a given solution can be neither stable nor unstable, if it is 
itself invariant under the involution $S$. However, this is not the case for 
the fixed point $R=1$ (in the absence of noise, i.e. for $\mu =1$) which is 
mapped by $S$ onto $R=-1$, 
so that we can only conclude that if $R=1$ is stable, then $R=-1$ must be 
unstable (as it is indeed the case). In other words, positive values of $R$ 
are characterized by a contracting dynamics towards $R=1$, while negative 
values depart from $-1$. If the multipliers are strictly positive, negative 
$R$-values cannot be asymptotically observed as they lie in the repelling 
part of the phase-space and the previous treatment in terms of a Langeving 
equation with an attracting force makes perfectly sense. On the other hand, 
if the multipliers can assume both signs, the dynamical rule allows visiting 
interchangeably the positive as well as the negative region. In principle, it 
is still possible to have, on the average, a global contraction provided that 
a longer time is spent in the positive region. Actually, this is the 
assumption more or less implicitely made in Ref.~\cite{tlpr}, where it 
was conjectured that no qualitative changes are expected when positive and 
negative multipliers come into play except for the degenerate case $p=q=1/2$. 
We see below that even if the scaling behaviour in the limit of vanishing
coupling is unaffected, this is not true and it indeed requires introducing
a different scaling parameter.

The most effective way we have found to analyse mapping (\ref{eq:Rmaps})
is by exploiting another property: the possibility to transfer the change of
sign of $\mu$ to $\gamma$. With this trick, the change of sign in 
Eq.~(\ref{eq:Rmaps}) can be effectively treated perturbatively being 
$\gamma$ a small parameter. 
More precisely, if $\mu(t)$ happens to be negative, we can 
assume it to remain nevertheless positive and perform the next iteration 
with $-R(t+1)$. It can then be seen that the resulting expression is the same 
as the original one after changing the sign of $\gamma$ and of $\mu(t+1)$. 
Now, irrespective of the sign of $-\mu(t+1)$, we assume it to be positive and 
transfer its sign to the next value of $\gamma$. In other words, we can 
iterate the mapping
\be
R(t+1) \; = \; |\mu(t)| \; \frac{R(t) + \gamma(t)}{1 + \gamma(t) R(t)} ,
\label{eq:Rmaps2}
\ee
where the sign of $\gamma(t)$ is that of $\prod_{s=1}^{t-1} \mu(s)$.
We can immediately see that even if $\mu$ is on the average more
positive than negative (or vice versa), the sign of $\gamma$ has no
preference, since it simply depends on the parity of the number of sign
changes. It is because of this reason that fluctuating multipliers
are qualitatively different from strictly positive ones: even an asymmetry
in the signs (a preference, say, for the positive values) implies that
the unstable and stable region (positive and negative values of $R$ in
the initial representation) are equally visited. 

The dichotomic structure of the noise $\gamma(t)$ allows expressing the 
stochastic map as the sum of a net drift plus a zero-average fluctuating term. 
Indeed, 
by calling $F_+(R)$ and $F_-(R)$ the l.h.s. of Eq.~(\ref{eq:Rmaps2}) whenever
$\gamma$ is positive, respectively, negative, we can write
\be
R(t+1) = \frac{1}{2}\bigg\{F_{+}(R(t)) + F_{-}(R(t))\bigg\} + 
\frac{\delta(t)}{2}\bigg\{F_{+}(R(t)) - F_{-}(R(t))\bigg\}
\ee
where $\delta(t)$ is again a dichotomic noise with entries equal to $\pm 1$.
More specifically, we obtain
\be
R(t+1) = \frac{|\mu|}{2}\frac{(1-\gamma^2)R}{1-\gamma^2 R^2} + 
|\mu| \delta(t) \frac{\gamma(1-R^2)}{1-\gamma^2 R^2}  .
\label{eq:Rmaps3}
\ee
In order to obtain an analytic expression for the probability density of
$R$, it is convenient to turn this equation into a continuous-time model. 

This is possible at the expense of assuming that the modulus 
fluctuations of $\mu$ are small, i.e. be writing $|\mu| = 1 + \nu$, and 
by then expanding the r.h.s. in (\ref{eq:Rmaps3}). By retaining terms up to
second order, we obtain
\be
\dot{R} = \gamma^2 R(R^2-1) + (\nu + \sol^2/2) R + \delta(t) \nu
 \gamma (1-R^2) ,
\label{eq:Rdot}
\ee
where $\sol^2$ is the variance of $\nu$.

Before going on, it is important to notice that the above equation must be
interpreted in the Ito sense as it arises from a discrete time stochastic
process with $\delta$-correlated noise \cite{Gard} (this problem did not 
arise in the previous section, since we were dealing with an additive 
noise). Moreover, it is instructive to notice that
the drift term in Eq.~(\ref{eq:Rdot}) is purely induced by noise: it arises
from the inhomogeneity of the stochastic process. This represents a direct
confirmation that contraction and expansions processes tend to compensate
each other as already anticipated in the beginning of this Section.

If the time variable is rescaled by a factor $\sol^2$, it is
immediately recognized that the dynamics of $R$ depends on just one
parameter
\be
  G = \frac{\gamma}{\sol} ,
\label{eq:scal2}
\ee
which is again a ratio between coupling strength and multiplier
fluctuations. However, there is an important difference with the parameter
$g = \gamma/\sigma^2$ introduced in the previous case, as it is seen by
noticing that in the small $\sol$ limit, the equality
\be
   \sol = \sqrt{2} \sigma
\ee
holds. Apart from the irrelevant numerical factor, it turns out that, in the
general case, the r.m.s. rather than the standard deviation enters as a
measure of multiplier fluctuations.

The Fokker-Planck equation corresponding to the Langevin process
(\ref{eq:Rdot}) in rescaled time units reads as
\be
\frac{\de P}{\de t} = -\frac{\de}{\de R} (A P) + 
\frac{1}{2}\frac{\de^2}{\de R^2} (B P)
\ee
where 
\be
A(R) = -G^2 R(1-R^2) + R/2
\ee
is the drift term, while
\be
B(R) = R^2 + G^2 (1-R^2)^2
\ee
is the diffusion coefficient. Since $4A(R) = dB/dR$, the stationary solution
is
\be
P(R)  = {N(G) \over \sqrt{G^2 (1 - R^2)^2 + R^2}}
\label{eq:pofr}
\ee
where $N(G)$ is the normalization constant discussed in Appendix B. 

An expression for the Lyapunov exponent can be obtained from
Eq.~(\ref{eq:shift}), by integrating over the above determined probability
distribution. Unfortunately, there is a crucial difference with the previous
case: we cannot simply expand the logarithm in powers of $\gamma$, since
this leads to computing the first moment of $P(R)$ which is already a
diverging quantity ($P(R)$ decays to zero as slowly as $1/R^2$). Obviously,
this is only a numerical artifact: the average of the logarithm itself is
still well defined and has a finite value.

Nevertheless, this is an indication that we must be much more cautious in
performing power expansions. In particular, this difficulty prevents
obtaining a general analytical expression analogous to that one obtained in
the previous section in terms of modified Bessel functions. In this case,
even obtaining an expression in the limiting case of small $G$ requires
a rather laborious work. In Appendix B we show that one can eventually
show that the non-trivial deviation with respect to the uncoupled limit is
given by
\be
\delta \Lambda = \frac{3 \sigma^2}{2\ln (1/G)} ,
\label{eq:dcle_pm}
\ee
Therefore, we see that also in the general case of positive/negative signs,
the leading dependence on $\varepsilon$ is of the type $1/\ln \varepsilon$,
as numerically observed. What is different is the dependence on the
multiplier fluctuations as testified by the presence of the parameter $G$
rather than $g$.

\subsection{Numerical results}

The first meaningful test of the analytical approach devised in the previous
sub-section concerns the probability distribution $P(R)$. In
Fig.~\ref{fig:histor} we report the outcome of a numerical experiment in
doubly-logairthmic scales (see the full dots). This allows seeing a
crossover from an initial decay as $1/R$ to the asymptotic decay $1/R^2$,
which represents the first qualitative confirmation of the thereotical
predictions. However, the agreement with expression (\ref{eq:pofr})
(represented by the dashed line) is more than just qualitative. In fact,
besides noticing the almost perfect overlap, one should also remember that
the only parameter entering Eq.~(\ref{eq:pofr}), i.e. $G$, has not been
fitted, but independently determined from the fluctuations of the local
multipliers. As a last remark, we would like to point out that the good
agreement is not totally obvious a priori at least for the reason that
the reduction from a set of coupled stochastic equations to a single
equation is not completely under control.

Moreover, it is instructive to compare the shape of this distribution
with the results predicted by the theory for strictly positive multipliers.
By expressing the probability density of Eq.~(\ref{eq:pofw}) in terms of
$R$, we find an exponential tail ($P(R) \simeq \exp(-gR)/R$). The power
law observed in Fig.~\ref{fig:histor} is also, therefore, an evidence of
a clear difference between the two regimes.

Finally, let us look at the deviations of the MLE plotted versus the
scaling parameter $G$. The data reported in Fig.~\ref{fig:scale_pm} have
been obtained for different noise amplitudes and either $\varepsilon=10^{-3}$
(diamonds) or $\varepsilon = 10^{-5}$ (circles). It is clearly seen that,
$\delta\Lambda\ln G/\sigma^2$ is constant, independently of the value
of $G$. This confirms the scaling behaviour predicted by 
Eq.~(\ref{eq:dcle_pm}). A more quantitative check can be made by comparing
the actual value of $\delta \Lambda \ln G/\sigma^2$  (about $1.1\approx 1.2$
in direct simulations) with the theoretical prediction (1.5)~. We believe
that the deviation is to be ascribed to the approximation made in reducing
the set of coupled stochastic equations to a single Langevin equation.

\section{Conclusions and perspectives}

In this paper we have developed a theoretical method that is able to explain
not only the scaling behaviour of the maximum Lyapunov exponent for a CML
in the small coupling limit, but provides a quantitative estimation
of its deviations from the uncoupled case. This was still lacking even
in the relatively simple case of strictly positive multipliers. However,
we have gone further, developing a treatment also for multipliers with
random signs. In both cases we have found that the correction to the
maximum Lyapunov exponent induced by local interactions actually depends
on a single scaling parameter which is nothing but the coupling strength
rescaled by the ``amplitude'' of multiplier fluctuations. However, the
scaling parameter is significantly different in the two cases: for
strictly positive multipliers, the fluctuation ``amplitude'' is the mean
square deviation $\sigma^2$ (see the definition of $g$ - Eq.~(\ref{defg})),
while in the case of random signs, the ``amplitude'' is the r.m.s. 
deviation (see the definition of $G$ - Eq.~(\ref{eq:scal2})).

Important differences can also be detected in the probability distribution
of the local ratios $R_i$ of the perturbation amplitude in two adjacent
sites: in the case of fluctuating signs there are long tails characterized 
by a power law decay.

Among the still open problems, there is certainly the exigence of a more 
rigorous procedure to solve the set of coupled Langeving equations. In fact, 
while the derivation of the set of coupled stochastic equations is the result 
of a well controlled perturbative approach, its reduction to a single
equation is based on a mean-field approximation whose validity cannot
be controlled a priori but only checked a posteriori.

Finally, we want to mention the possibility of extending this approach
to the case of weakly coupled attractors, where time is continuous from
the very beginning. This is certainly the most stimulating perspective that
is also supported by the preliminary observation that the Langevin equation
(\ref{eq:wappr}) is obtaind also in the case of two weakly coupled 
differential equations \cite{pikoco}.

\vspace{1.0cm}
{\bf Acknowledgments} We thank the Institute of Scientific Interchange 
(I.S.I.) in Torino, Italy, where this work was started and partly carried on.
We are also indebted to J.M. Parrondo for having pointed out the opportunity
to adopt the Ito interpretation.

\appendix \section{Linear limit}

This Appendix is devoted to the analysis of Eq.~(\ref{eq:wdot}) in the
linear limit,
\be
\dot w_i = -\gamma (w_{i+1} - 2 w_i + w_{i-1}) + \psi_i(t) - \psi_{i-1}(t) ,
\label{eq:wlinear}
\ee
where we have introduced $\psi_{k} = \ln\{m_k(t)\}$. This is apparently a 
discretized Edwards-Wilkinson equation \cite{basta}, but the spatial 
structure of the noise prevents the onset of any roughening phenomenon
(as commented in the main body of the paper).

To solve this equation, it is convenient to perform a spatial Fourier
transform, since it leads to a set of uncoupled equations,
\be
\dot w(k,t) = -2 \gamma (1- \cos (k)) w(k,t) + (1-{\rm e}^{ik}) \psi(k,t),
\label{eq:fou}
\ee
where $w(k,t)$ is a complex number that can be decomposed into a real
and imaginary part ($w(k,t) = x(k,t) + i\;y(k,t)$), which satisfy the
same equation
\be
\dot x(k,t) = -2 \gamma (1-\cos (k)) x(k,t) + \eta(k,t),
\label{eq:lang2}
\ee
where the noise term $\eta$ is $\delta$-correlated,
\be
\langle \eta(t)\eta(t') \rangle = 2 \sigma^2\,(1-\cos (k))\delta(t-t') .
\ee
Accordingly, all Fourier modes obey the same Gaussian distribution function,
\be
P\{w\} \sim  \exp\bigg(-\frac{\gamma (x^2 + y^2)}{2\sigma^2}\bigg).
\ee
The probability distribution of $w_i$ on a single site is easily obtained by
summing the independent distributions corresponding to all modes. As a
result, the distribution of $w_i$ is also Gaussian and its variance is
$\sigma^2$, as if we had neglected the spatial coupling in
Eq.~(\ref{eq:wlinear}).

\section{Lyapunov correction}

In this Appendix we determine the nontrivial part of the leading correction
to the MLE in the general case. We start by computing the normalization
constant. It is convenient to exploit the invariance of $P(R)$ under parity
change and the transformation $R \to 1/R$, to express the normalization
condition as
\be
1 = 4 \int_{0}^{1} dR P(R) = 4 N(G)  \int_{0}^{1} 
  \frac{dR}{\sqrt{R^2 + G^2(1- R^2)^2}} .
\label{eq:intnor}
\ee
Since an explicit analytical expression for the above integral does not
exist, we shall limit ourselves to studying the small-$G$ limit. One
cannot simply expand the denominator, as it gives rise to a non-integrable 
singularity in $R=0$. It is, instead, convenient to introduce the variable 
$x = R/G$. Afterwards, one can expand the integrand in powers of $G$ without 
encountering undesired divergences. By retaining the leading terms, we find
\be
\frac{1}{N} \simeq  4 \int_{0}^{1/G} \frac{dx}{\sqrt{1 + x^2}}  \simeq
4 \ln\bigg(1/G\bigg)
\ee 
From Eqs.~(\ref{eq:lyap},\ref{eq:tgratio}), it turns out that the estimation
of $\delta \Lambda$ requires computing the mean value of
$L(R_1,R_2) \equiv \ln |1 + \gamma R_1 + \gamma/R_2|$, i.e.
\be
\delta \Lambda = \int_{-\infty}^{\infty} \int_{-\infty}^{\infty} dR_1 dR_2 
P(R_1) P(R_2) L(R_1,R_2) 
\ee
Thanks to the equality $L(R_1,R_2) = L(1/R_2,1/R_1)$ and to the invariance
of $P(R)$ under the transformation $R \to 1/R$, we can write the Lyapunov
correction as the sum of three different contributions, namely
\BE
\delta \Lambda  & \equiv &  \delta_1 + \delta_2 + \delta_3 \\
 & = & \int_{-1}^{1} \int_{-1}^{1} dR_1 dR_2
P(R_1) P(R_2) \bigg\{L(R_1,1/R_2) + 2 L(R_1,R_2) +
L(1/R_1,R_2) \bigg\} \nonumber
\EE
with an obvious meaning of the new symbols.

In analogy with the computation of the normalization constant, we introduce
the variables $x = R_1/G$ and $y = R_2/G$. As a consequence, the expressions
for the three contributions write as
\BE
 \delta_1 & = & N^2  \int_{-1/G}^{1/G} \int_{-1/G}^{1/G} dx dy
 \frac{\ln| 1 + \gamma G x + \gamma G  y |} 
{\sqrt{x^2 + (1-G^2 x^2)^2}\sqrt{y^2 + (1-G^2 y^2)^2}} \\
  \delta_2 & = & 2 N^2  \int_{-1/G}^{1/G} \int_{1/G}^{1/G} dx dy
 \frac{\ln| 1 + \gamma G x + \gamma/(Gy) |}  
{\sqrt{x^2 + (1-G^2 x^2)^2}\sqrt{y^2 + (1-G^2 y^2)^2}} \\
 \delta_3 & = & N^2  \int_{-1/G}^{1/G} \int_{-1/G}^{1/G} dx dy       
\frac{\ln| 1 + \gamma/(Gx) + \gamma/(Gy) |}
{\sqrt{x^2 + (1-G^2 x^2)^2}\sqrt{y^2 + (1-G^2 y^2)^2}}
\EE
The inequalities $\gamma \ll G \ll 1$ imply that the contributions
proportional to $\gamma G$ in the arguments of the logarithms can be
neglected so that $\delta_1$ is negligible altogether. Moreover,
$G$ can be always neglected in the denominators, so that the leading
contribution to the MLE can be determined by just estimating the two
integrals
\BE
 \delta_2 & = &  2 N \int_{-1/G}^{1/G} dy
 \frac{\ln| 1 + \sol/y |}{\sqrt{1 + y^2}}
\label{eq:del2}\\
 \delta_3 & = &  N^2 \int_{-1/G}^{1/G}
 \int_{-1/G}^{1/G} dx dy
 \frac{\ln|1 + \sol/x + \sol/y |}
  {\sqrt{1 + x^2}\sqrt{1 + y^2}}
\label{eq:contr-c}
\EE
where we have re-introduced the parameter $\sol$ for later convenience.
We start discussing $\delta_2$; it cannot be computed by expanding the
logarithm as this leads to an unphysical divergence. It is, instead,
helpful to split this contribution into two parts
\BE
\delta_2 &=& \delta'_2 + \delta_2'' \\
& = & 2N \int_{-1/G}^{1/G} dy \frac {\ln |y + \sol|}{\sqrt{1+y^2}}
 - 2N \int_{-1/G}^{1/G} dy \frac {\ln |y|}{\sqrt{1+y^2}} \nonumber
\EE
The first integral can be estimated by introducing the variable
$w = y + \sol$ and thereby expanding the denominator in powers of
$\sol$. By retaining terms up to the second order, we find that
$\delta'_2$ can be written as
\be
 \delta'_2  =  2 N \int_{-1/G+\sol}^{1/G+\sol} dw
 \left \{ \frac{\ln |w|}{\sqrt{1+w^2}} +
 \sol \frac{w \ln |w|}{(1+w^2)^{3/2}} +
  \frac{\sol^2}{2} \frac{w^2 \ln |w|}{(1+w^2)^{5/2}}(2w^2 -1)
  \right \}
\ee
By expanding the zeroth-order term around the integral boundaries in powers
of $\sol$, we find that it is equal to $-\delta_2''$ plus corrections of
the order $\gamma^2 \ln G$. A contribution of the same order is obtained
also by integrating the linear term in $\sol$. However, the leading
contribution to the MLE comes from the second-order term which, in the
small-$G$ limit can be written as
\be
\delta_2 = 2N \sol^2 \int_0^\infty dw
  \frac{\ln w}{(1+w^2)^{5/2}}(2w^2-1)  .
\ee
The integral can be analytically solved and turns out to be equal to 1, so
that \cite{notel}
\be
\delta_2 = 2N(G) \sol^2 = \frac{\sol^2}{2\ln (1/G)}
\label{del2f}
\ee
The determination of $\delta_3$ in principle requires even more cumbersome 
calculations, as it involves a double integral. However, formally deriving
the expression for $\delta_3/N^2$ with respect to $G$, we find that, up
negligible corrections,
\be
\frac{d\delta_3/N^2}{dG}= -\frac{4}{G} \int_{-1/G}^{1/G} dy
    \frac{\ln | 1 + \sol/y |}{\sqrt{1+y^2}} .
\ee
As the integral in this expression is exactly the same involved in the
definition of $\delta_2$ (see Eq.~(\ref{eq:del2})), we can write
\be
\frac{d\delta_3/N^2}{dG}= -\frac{2\delta_2}{GN}
\ee
Upon substituting the expression for $\delta_2$ (see Eq.~(\ref{del2f}),
the above equations can be rewritten as
\be
\frac{d\delta_3/N^2}{dG}= -\frac{4\sol^2}{G} \quad,
\ee
which, after integration, yields
\be
  \delta_3 = \frac{\sol^2}{4\ln (1/G)}
\ee
In conclusion, we find that
\be
\delta \Lambda = \frac{3 \sigma^2}{2\ln (1/G)} \quad,
\ee
where we have preferred to introduce the explicit dependence on the physical
parameter $\sigma$ rather than $\sol$.



\begin{figure}
\caption{Probability distribution of $w$ for two different values of $g$:
2.4 (a) and
0.021 (b). In both cases, circles refer to the numerical histograms, obtained 
by iterating Eq.~(\pr\ref{eq:cmlt}) for $5\times10^7$ time-steps
(discarding the first $10^3$ iterations) on a lattice of $L=300$ sites. The
solid curves correspond to the analytic formula (Eq.~(\pr\ref{eq:pofw}) with
$\alpha = 1$). The dashed curve in (b) is obtained by fitting $\alpha$
which is estimated to be equal to 0.75.}
\label{fig:histow}
\end{figure}

\begin{figure}
\caption{The Lyapunov exponent versus the scaling parameter $g$ in the case
of strictly positive multipliers. The data is obtained by varying $\eps$, 
$\sigma$ and for both choices of the probability distribution of the local 
multipliers (the cases A and B discussed in the text). In a), the shift
$\delta \lambda$ defined in Eq.~(\pr\ref{eq:shift}) is reported, while in b)
the total shift $\Delta \Lambda$ (see Eq.~(\pr\ref{eq:globly})) is plotted.
The solid curves correspond to the analytic expression. The dashed line in
b) is the analytical curve arbitrarily shifted to show that a
``renormalization''  of the scaling parameter could account for the
remaining discrepancy with numerical data.}
\label{fig:scale}
\end{figure}

\begin{figure}
\caption{Log-log plot of the probability distribution $P(R)$ to highlight 
the power-law behaviour. Circles, triangles and diamonds refer to 
$G=\sqrt{6}\cdot 10^{-2}, \sqrt{6}\cdot 10^{-3}, \sqrt{6}\cdot 10^{-5}$,
respectively. The simulation details are as in Fig.~\pr\ref{fig:histow}.
The various curves represent the analytical results as from
Eq.~(\pr\ref{eq:pofr}).}
\label{fig:histor}
\end{figure}

\begin{figure}
\caption{The Lyapunov exponent versus the scaling parameter $G$ in the case
of fluctuating multipliers. The Lyapunov correction $\delta \Lambda$ is
normalized so as to emphasize the $1/|\ln G|$ dependence. Circles correspond
to $\eps = 10^{-5}$ while diamonds to $\eps=10^{-3}$. The straight solid line 
represents the theoretical result (\pr\ref{eq:dcle_pm}).}
\label{fig:scale_pm}
\end{figure}


\begin{thebibliography}{99}
\bibitem{notel} We have solved it with the help of Maple software package.
\end{thebibliography}
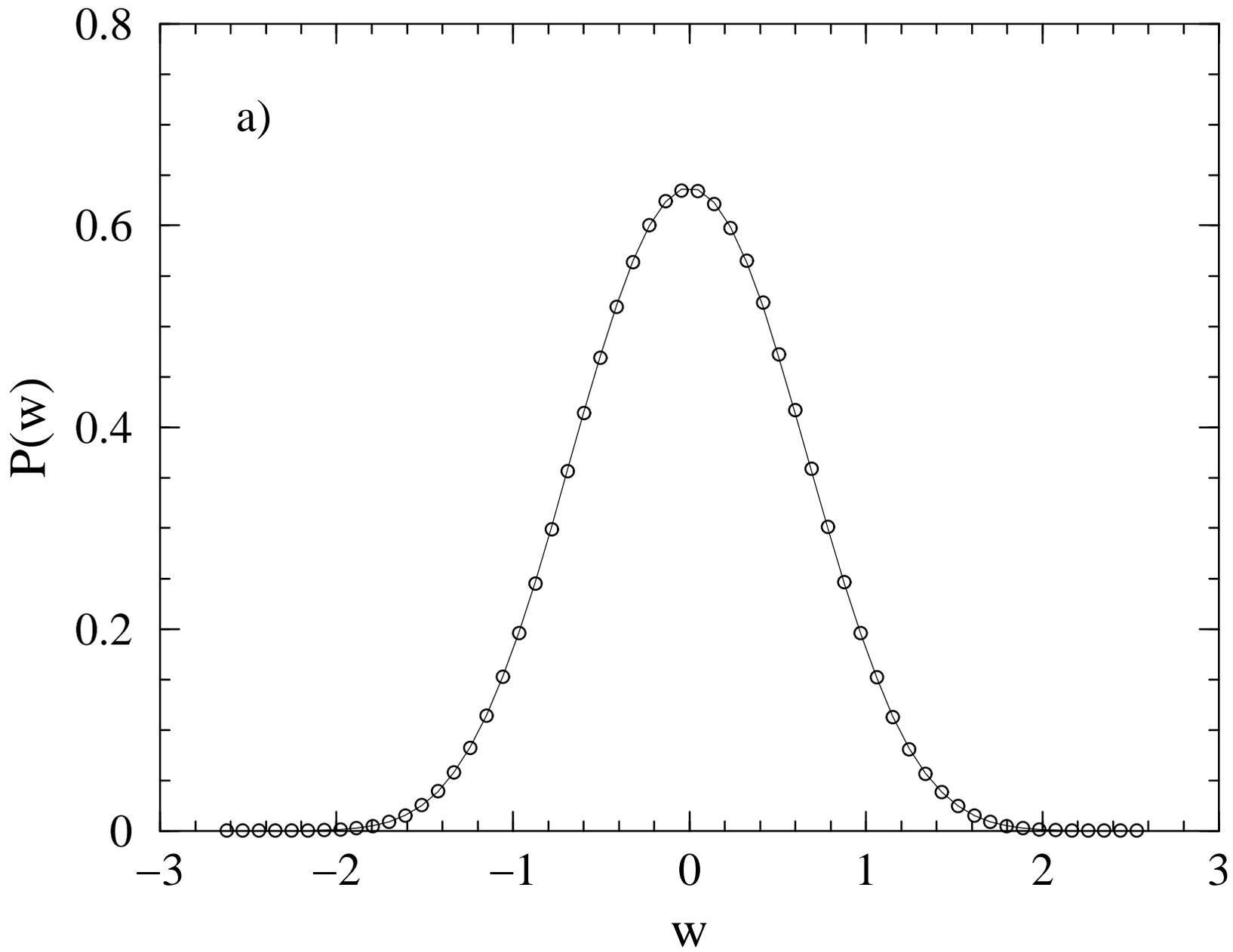
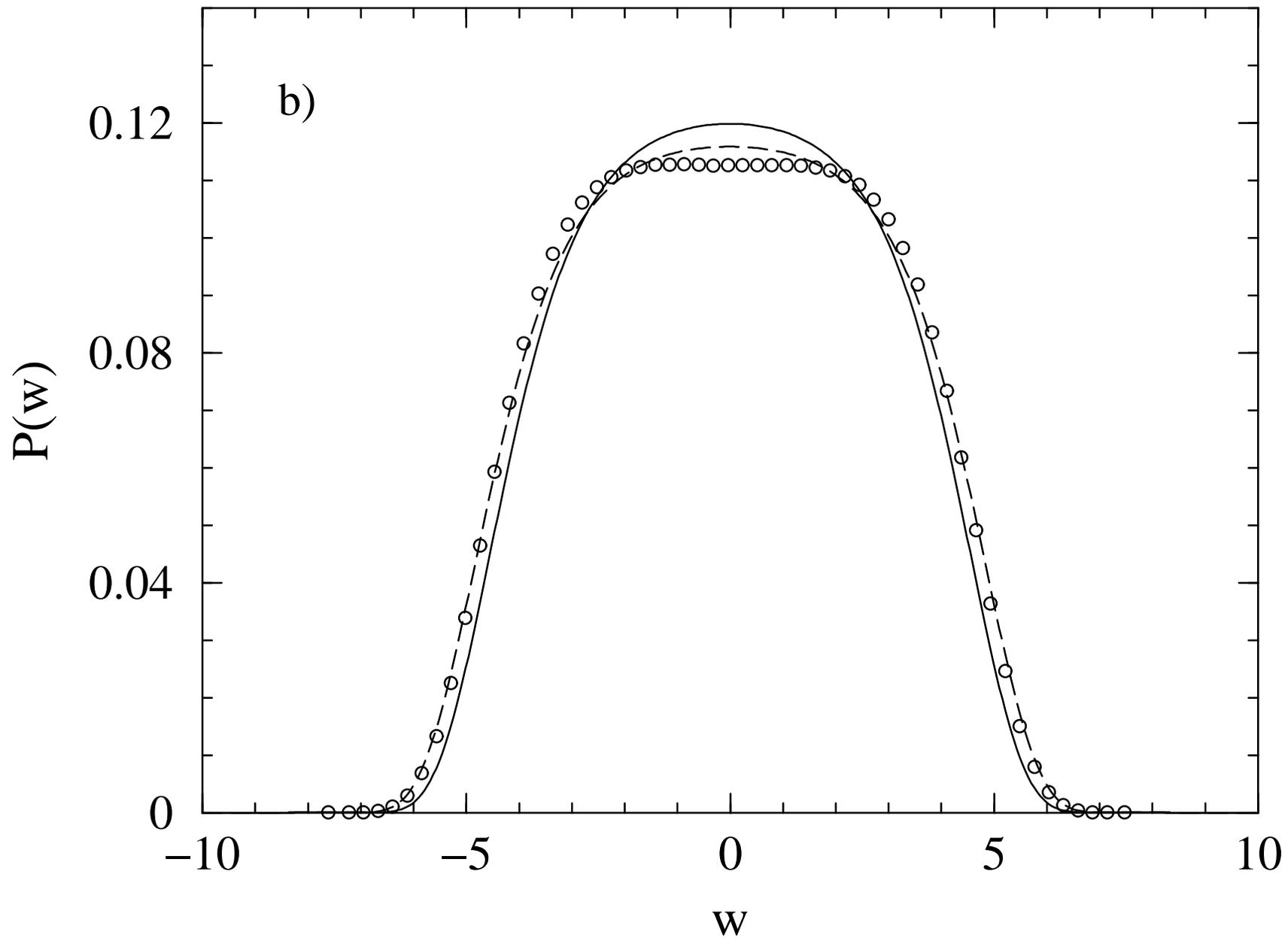
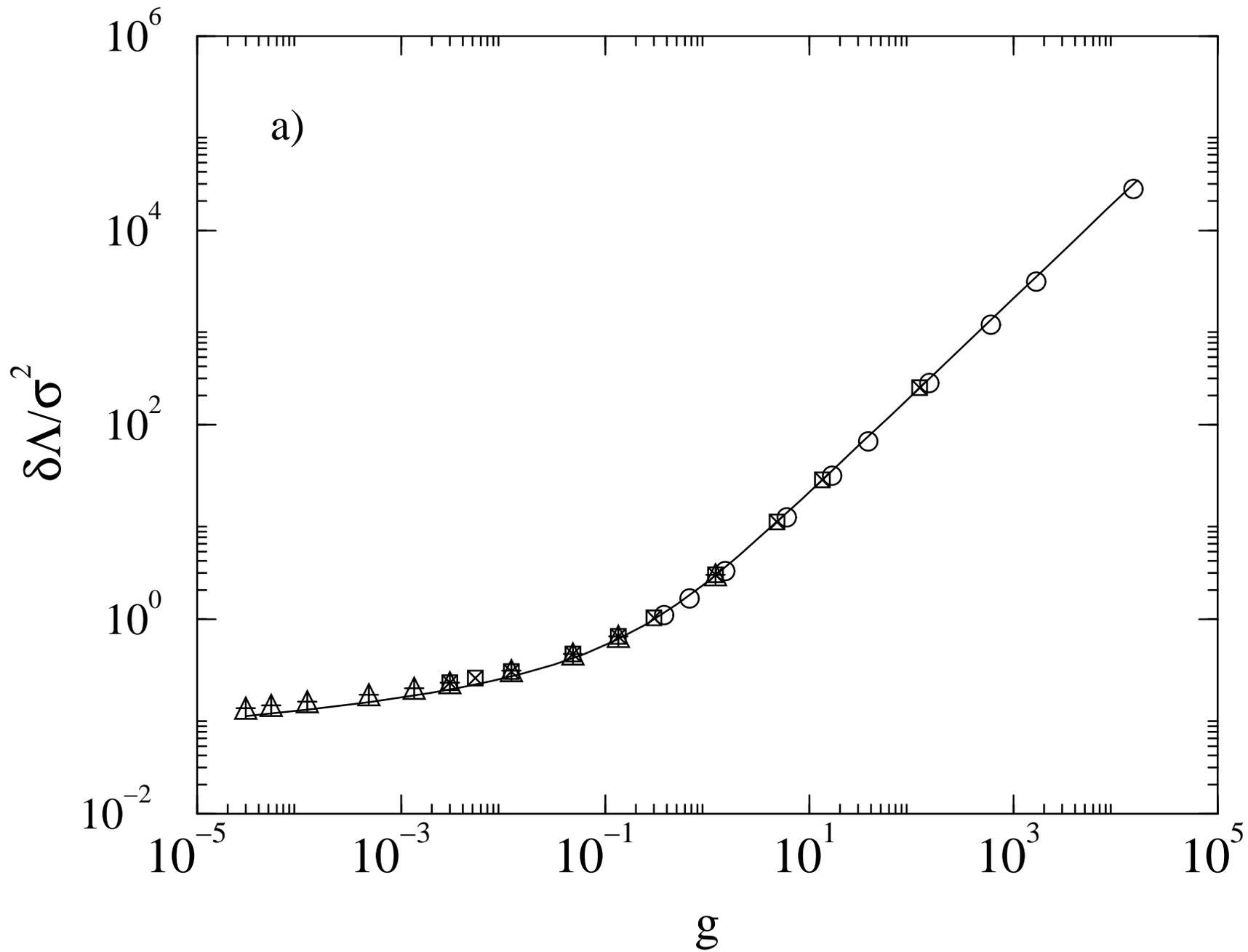
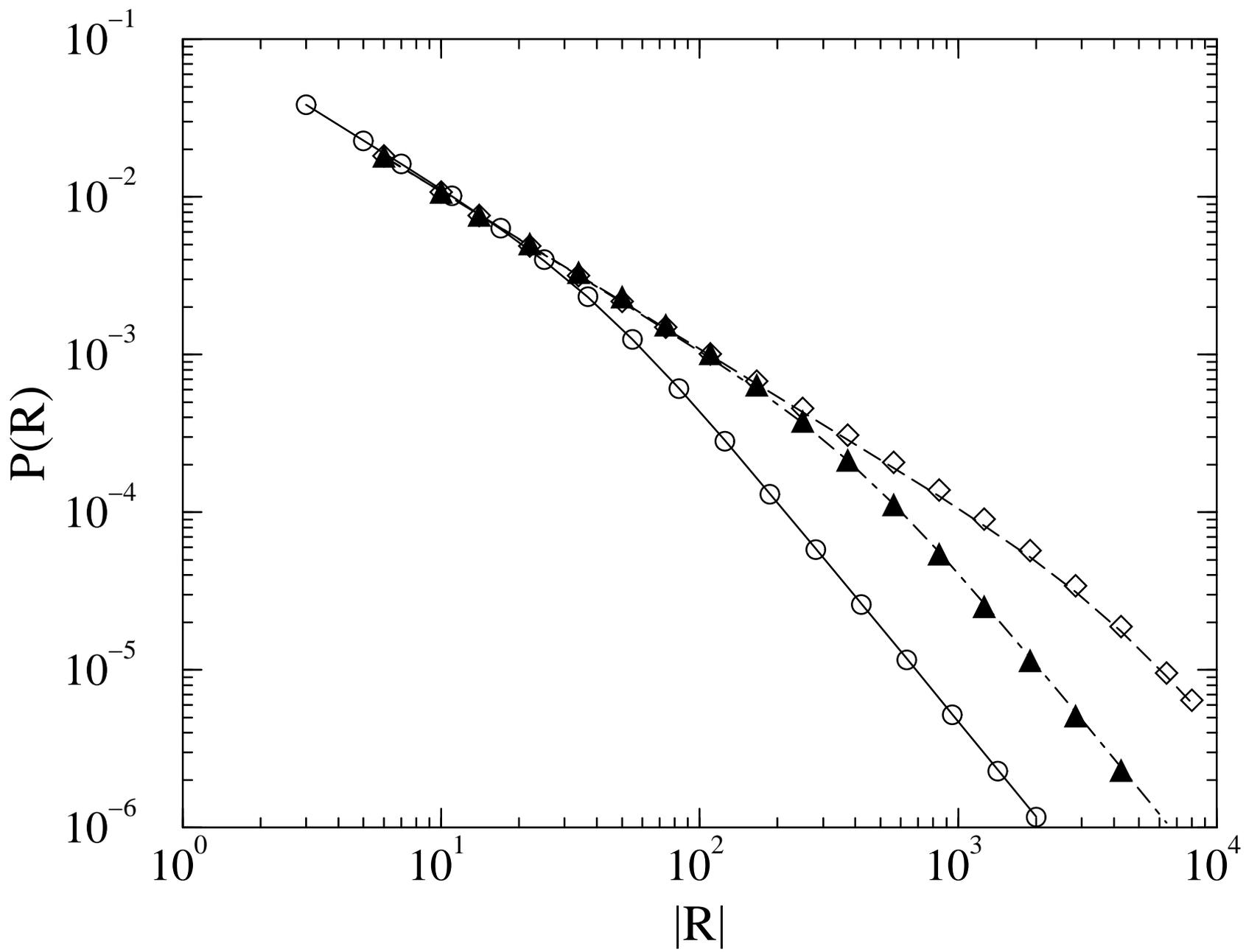
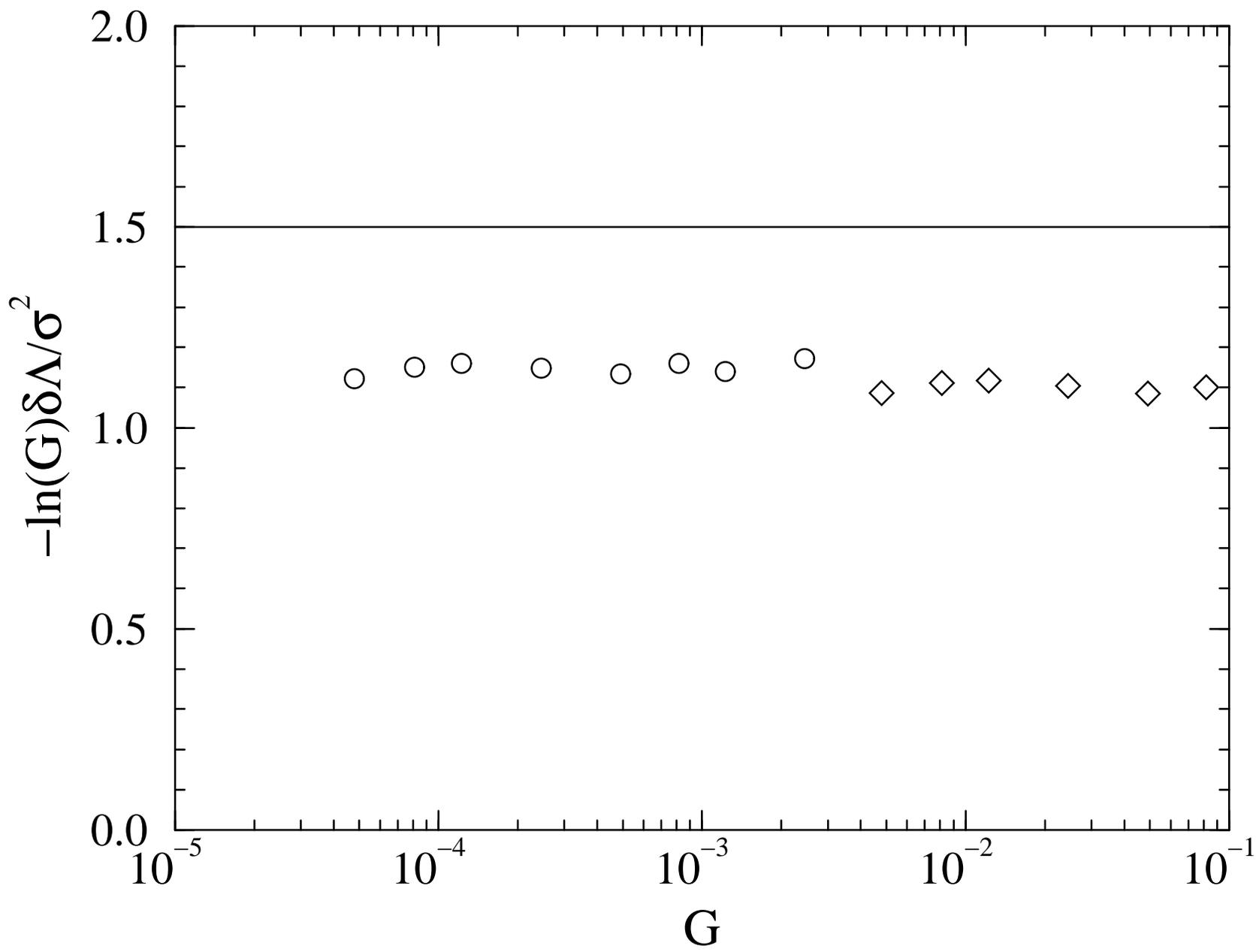
\end{document}